\documentclass[12pt, a4paper]{article}

\usepackage[a4paper, top=2.5cm, bottom=2.5cm, left=2.5cm, right=2.5cm]{geometry}
\usepackage[T1]{fontenc}
\usepackage[utf8]{inputenc}
\usepackage{lmodern}
\usepackage[english]{babel}
\usepackage{graphicx}
\usepackage{booktabs}
\usepackage{amsmath}
\usepackage{amssymb}
\usepackage{setspace}
\usepackage[authoryear, round]{natbib}
\usepackage{authblk}
\usepackage{lineno}
\usepackage{hyperref}
\usepackage{caption}
\usepackage{float}
\usepackage{titlesec}
\usepackage{multirow}
\usepackage{colortbl}
\usepackage{tikz}
\usetikzlibrary{shapes.geometric, shapes.symbols, arrows.meta, positioning, fit, calc, backgrounds}

\usepackage{xcolor}
\usepackage[normalem]{ulem}
\definecolor{revred}{RGB}{178,24,43}

\newcommand{\redsout}{\bgroup\markoverwith{\textcolor{revred}{\rule[0.5ex]{2pt}{0.7pt}}}\ULon}

\newcommand{\glyph}[1]{\raisebox{-0.25ex}{\includegraphics[scale=0.83]{glyphs/#1}}}

\hypersetup{
    colorlinks=true,
    linkcolor=blue,
    filecolor=magenta,      
    urlcolor=cyan,
    citecolor=blue,
}

\title{\textbf{Steelmakers increasingly cite scrap quality \\ when closing material loops \\ with their customers}}

\author[1,2,3,4,*]{Peter Klimek}
\author[1,*]{Jaber Fooladi}

\affil[1]{Supply Chain Intelligence Institute Austria (ASCII), Metternichgasse 8, A-1030 Vienna, Austria}
\affil[2]{Section for the Science of Complex Systems, CeMSIIS, Medical University of Vienna, Vienna, Austria}
\affil[3]{Complexity Science Hub Vienna, Vienna, Austria}
\affil[4]{Division of Insurance Medicine, Karolinska Institutet, Department of Clinical Neuroscience, Karolinska}
\affil[*]{\small \texttt{\{peter.klimek, jaber.fooladi\}@ascii.ac.at}}
\date{\today}

\begin{document}

\maketitle

\begin{abstract}
The transition to a circular economy is pivotal for industrial decarbonization, particularly in the energy-intensive steel sector. While recycling scrap via electric arc furnaces offers a low-carbon alternative to primary production, the accumulation of contaminants in post-consumer steel threatens to render secondary material unusable for high-grade applications. This quality problem creates a market failure where traditional spot markets cannot guarantee the material provenance required for high quality (e.g. automotive-grade) steel. Here we show, using a massive longitudinal analysis of 1,163,561,858 news articles combined with global trade network data, that the industry addresses this partly through direct, vertically integrated alliances. We demonstrate that steelmakers are bypassing intermediaries to forge sovereign, closed-loop ties with manufacturers, effectively substituting market mechanisms with organizational integration. This reorganization coexists with a division between circular network cores in the Americas and Europe that pair their steel and scrap flows most tightly within their region, and economies whose steel and scrap flows are far less matched. At firm level, the quality of the returned scrap is a stated reason in twelve of thirty-six reciprocal loops closing after 2020 and in none of the eight loops detected before. Our results indicate that the circular economy is evolving into a geopolitical contest for material control, where competitive advantage is defined by the sovereign possession of closed material cycles rather than mere cost efficiency. 
\end{abstract}

\newpage

\section{Introduction}

For over a century, the global steel industry has operated on a linear "take-make-waste" model, predicated on the extraction of virgin iron ore. We are now transitioning into what has been called the steel scrap age \cite{pauliuk2013steel}, a pivotal era where the global stock of steel embedded in buildings, vehicles, and appliances has become sufficiently large that recycling is poised to overtake mining as the primary source of production. This transition signifies more than a shift in raw materials; it represents a fundamental restructuring of global industrial metabolism, shifting the value creation paradigm from resource extraction to material regeneration.

This transformation is a core element of the sector's strategy to reduce its emissions. The traditional blast furnace-basic oxygen furnace (BF-BOF) route is carbon-intensive, relying heavily on coal to reduce iron ore. In contrast, the electric arc furnace (EAF) route, which melts scrap (or directly reduced iron, which cuts emissions without closing a loop) using electricity, represents the "low-hanging fruit" for meeting Paris Agreement targets \citep{cullen2012}. As developing nations complete their massive infrastructure build-outs, their in-use stock of steel will eventually mature and return to the system, transforming ferrous scrap from a generic waste byproduct into a strategic asset.

However, this theoretical concept faces a severe practical and physical constraint: the quality problem. Modern manufacturing inextricably links steel with other materials; for instance, when end-of-life vehicles (ELVs) are shredded, copper wiring and sensors become commingled with the ferrous fraction. These tramp elements accumulate from one product generation to the next, causing hot shortness (cracking) during rolling and threatening to render the global steel stock unsuitable for high-grade applications (\cite{nakamura2017, daehn2017}). This creates a "market for lemons" market failure \citep{akerlof1970}, where the scrap spot market fails to signal quality, and steelmakers can no longer rely on anonymous spot transactions to secure inputs of sufficient purity.

To mitigate this, companies are pursuing two distinct strategies: technological solutions such as advanced sensor-based sorting, and organizational solutions involving direct buy-back schemes and closed-loop supply chain integration. While sorting technologies are advancing, a critical maturity gap remains. Direct buy-back schemes are currently well-established only for pre-consumer (post-industrial) scrap, such as stamping off-cuts, which are chemically predictable and high-value. In contrast, post-consumer loops, dealing with the complex, contaminated flows from end-of-life products, remain the critical frontier of industrial organization.

The scarcity of high-quality scrap transforms it into a strategic asset, leading to a ``geopolitics of scrap.'' Concerns have been raised regarding a looming era of resource nationalism, where mature economies with saturated stocks seek to retain secondary materials for domestic decarbonization \citep{charpin2021}. This has raised further concerns about a potential North-South divide, where industrialized nations enact protectionist barriers to secure their sovereign circularity, leaving developing nations in the accumulation phase reliant on primary extraction or low-quality waste imports \citep{gregson2015, nechifor2020}. Such developments could aggravate the quality problem by contaminating the more pristine scrap stock of developing countries with polluted scrap from other regions. However, empirical evidence for this structural transformation remains sparse. While the intent to close loops is well-documented in policy papers and material flow analyses, it is largely unknown whether the global steel trade network has actually begun to reorganize in response to these quality and geopolitical pressures, or if it remains tethered to traditional linear market dynamics.

In this paper, we investigate whether the industry is strategically shifting from interest to execution. While the necessity of high-quality recycling is well-understood, we address the \textit{how} and \textit{when} of its implementation. We employ a multi-scale quantitative research design to capture this transition. First, we utilize a massive longitudinal news mining approach on 1,163,561,858 articles to identify strategic alliances and direct relationships, and whether they signal pre- or post-consumer loops. Second, we complement this with a macro-level global trade network analysis, to quantify the extent of these reciprocal circular relationships physically. Our contribution provides empirical evidence that the steel industry is bridging the gap between technical necessity and industrial reorganization, moving towards a future defined by direct, verified, and reciprocal material loops.

\section{Literature Review and Hypothesis Development}

The transition from a linear ``take-make-waste'' model to a circular economy is one of the most significant structural changes in modern economic history. This section synthesizes distinct streams of literature, including the physical realities of material flow analysis (MFA) and the strategic reorganization of supply chains under resource constraints. Bringing together these disparate fields reveals a significant gap, namely the need for high-purity recycling in post-consumer loops and its effect on the topological restructuring of global trade networks into reciprocal configurations \citep{sun2024}.

\subsection{Global Stock Saturation and Trade Asymmetries}
Industrialized societies are predicated on steel, accumulating substantial in-use stocks that eventually serve as the predominant reservoir for future production. \citet{pauliuk2013} formalize this transition through the steel scrap age hypothesis, utilizing a dynamic stock model to demonstrate that per capita steel stocks eventually saturate in mature economies. As infrastructure expansion decelerates, the rate of scrap generation accelerates relative to demand. This physical saturation changes industrial structure: the primary input shifts from virgin iron ore to ferrous scrap, making future industrial capacity dependent on historical accumulation \citep{ref627}.

However, it has been argued that this transition may create a structural divergence in the global economy. Mature economies, characterized by high scrap availability and advanced EAF technology, face structural incentives to close their loops internally. To secure decarbonization pathways, these regions are driven to retain high-quality secondary resources domestically. In contrast, developing economies are largely in an accumulation phase. Rapid urbanization drives demand for primary steel that domestic scrap generation cannot yet meet \citep{ref629, cullen2012}, creating a risk of exogenous feedstock reliance where they depend on imported, often lower-quality scrap or remain locked into carbon-intensive primary production \citep{ref1034, ref1035}.

China plays a distinct role in this dynamic. The implementation of the ``National Sword'' policy and stringent scrap import bans may be interpreted not merely as isolationism, but as regulatory decoupling. By decoupling from the global trade in low-quality waste, China is effectively managing its domestic circulation and quality control, establishing an internal system of sovereign circularity \citep{wang2021}.

\subsection{The Quality Constraint: Impurity Accumulation and Market Inefficiency}
The circularity of steel is physically constrained by the accumulation of tramp elements. \citet{nakamura2012} demonstrated that copper contamination, in particular, induces hot shortness (cracking during hot rolling), rendering the resulting steel unsuitable for high-quality applications like automotive sheets. This metallurgical limit creates a bifurcation in the market between a downcycling loop and a high-quality closed loop.

Recent empirical benchmarks highlight the severity of this constraint. A 2025 study by \citet{imt2025} specifically assessed the feasibility of closed-loop recycling for automotive steel. They found that standard ``market scrap'' (fragmented and shredded flows) typically exceeds a copper content of 0.4\%. In stark contrast, high-grade automotive flat steel requires copper levels below 0.15\%. The study demonstrated that deep dismantling and closed-loop systems can achieve copper levels below \textbf{0.1\%}, meeting these stringent requirements.

This discrepancy exposes a fundamental market inefficiency. The spot market for scrap suffers from severe information asymmetry, constituting a classic ``market for lemons'' \citep{akerlof1970}. Because the chemical composition of shredded scrap is costly to verify before melting, the price mechanism fails to signal quality effectively. This leads to adverse selection, where high-quality buyers exit the spot market because they cannot distinguish between prime and contaminated material without incurring prohibitive transaction costs \citep{liu2018}.

\subsection{Organizational Integration: Vertical Coupling}
To resolve these constraints, the industry is shifting from technological reliance to organizational integration. While sensor-based sorting technologies are advancing, they currently lack the ability to guarantee the traceable provenance required for high-grade applications. Consequently, a distinction in loop maturity is emerging. \textit{Pre-consumer} loops, involving process scrap like stamping off-cuts, are mature and well-established due to their high purity and existing logistics \citep{hatayama2014tracking}. \textit{Post-consumer} loops, dealing for instance with ELVs, represent the emerging frontier where ownership transfer creates a significant barrier to quality assurance \citep{davis2024reuse, wagner2025environmental}.

To bridge this gap, firms are exploring vertical integration mechanisms. These include steel-as-a-service models, where steelmakers retain ownership of the material to ensure its return, and digital product passports to create an informational infrastructure that lowers verification costs. Most critically, \citet{daehn2017} argue that direct supply coupling is mandatory. To prevent global contamination, steelmakers must establish direct reciprocal ties with OEMs, bypassing the potentially contaminated pool of the open market.

\subsection{Hypothesis Development}
Based on these structural and strategic drivers, we propose two hypotheses regarding the evolution of the global steel production network:

\begin{quote}
\textbf{Hypothesis 1 (Macro-Structural):} As regional EAF capacity increases in mature economies, global trade networks will exhibit a topological shift toward bidirectional trade asymmetry, where steel exports and scrap imports increasingly overlap within sovereign jurisdictions.
\end{quote}

Because an arc furnace can also run on reduced iron, we test H1 as two claims: that steel exports and scrap imports are matched within regions beyond chance, and that this co-varies with EAF share. Sections 4.2 and 5 support the first but not the second.

\begin{quote}
\textbf{Hypothesis 2 (Micro-Strategic):} To mitigate quality uncertainty, producers of high-grade steel will preferentially form direct vertical alliances with OEMs, bypassing intermediary markets to secure high-purity pre- and post-consumer flows.
\end{quote}
\section{Data and Methods}

To test the hypotheses developed in the previous section, we employ a multi-scale network analysis design that integrates semantic and topological data layers. We begin with the micro-level analysis as a leading indicator of strategic intent, followed by the macro-level analysis to quantify the systemic shift in physical flows.

\subsection{Micro-Level News Mining (CC-News)}
While trade-data-based macro-analysis elucidates shifts in mass balance on a country level, it lacks the granularity to identify the specific firm-level alliances driving these flows. To study this topology, we implemented an AI-driven analysis of global industrial news.

\subsubsection{Data Source: The CC-News Corpus}
We utilized the \textbf{CC-News dataset} (Common Crawl News), a publicly accessible archive of news articles from global sources. In contrast to structured trade databases, CC-News offers unstructured, ``in-the-wild'' reporting, capturing press releases and industry journalism that reveal specific supplier-buyer contracts and strategic alliances that aggregated trade statistics frequently obscure \citep{tornberg2023}. The results use the archive completed to September 2026, containing 1,163,561,858 news items in 49 languages, as input to the analytic approach described in Section \ref{sec:pipeline}.

\subsubsection{Multilingual Extraction Pipeline}\label{sec:pipeline}
The analytic approach (the ``pipeline'' depicted in Figure \ref{fig:pipeline}) covers all 49 languages and runs in six stages. Appendix \ref{sec:pipeline_details} describes each stage in technical detail, and Appendix \ref{sec:corpus_examples} gives examples.
\begin{enumerate}
\item \textit{Corpus preparation.} The text of each archived news page is extracted and split by detected language, so that a record is one page in one language, dated by the month in which it was archived.
\item \textit{Vocabulary screen.} Each record is filed under its language and retained only if it carries at least one steel or scrap term from that language's word list\footnote{Steel and scrap terms, for example: English \textit{steel}, \textit{steelworks}, \textit{scrap}, \textit{ferrous scrap}; German \textit{Stahl}, \textit{Stahlwerk}, \textit{Schrott}, \textit{Stahlschrott}; Turkish \textit{\c{c}elik}, \textit{\c{c}elikhane}, \textit{hurda}, \textit{hurda demir}; Chinese \glyph{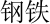} (steel) and \glyph{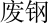} (steel scrap); Persian \glyph{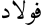} (steel) and \glyph{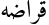} (scrap). The lists for all 49 languages are in the replication files.}.
\item \textit{Selection gate.} Three rules select 914,592 of the retained records for the language model: in languages small enough to be read in full every record is selected; elsewhere a record is selected when it names both steel and scrap, or when it names an unambiguous industry term together with scrap or uses a closed-loop phrase\footnote{Unambiguous industry terms, for example English \textit{steelworks}, \textit{ferrous scrap}; German \textit{Stahlwerk}, \textit{Stahlschrott}; French \textit{aci\'erie}, \textit{ferraille}. Closed-loop phrases, for example English \textit{closed loop}, \textit{take-back}, \textit{circular economy}, \textit{scrap return}; Turkish \textit{kapal\i\ d\"ong\"u} (closed loop) and \textit{hurda iadesi} (scrap return).}.
\item \textit{Relation extraction.} A language model reads each selected record, with one prompt for all languages, and returns each transactional relation (a transaction, supply agreement or take-back arrangement between two named firms involving steel or ferrous scrap) as supplier, buyer, item and direction of flow, with the sentence it rests on.
\item \textit{Entity resolution.} The spellings and scripts under which one firm appears are mapped to a single English name, a parent group and a home country, so that each firm is counted once.
\item \textit{Loop assembly and verification.} Relations are paired into candidate loops, a firm pair with steel flowing one way and ferrous scrap the other, matched on resolved name or parent group in either orientation; duplicate loops are merged, and every candidate loop was read against its full sources in the original language, and 44 of 56 survived.
\end{enumerate}
A loop is dated by the year it closes, the later of its two legs' first appearance, and is coded as quality-motivated when a source gives a property of the returned scrap (grade, cleanliness, residual content, composition, segregation) as a reason for the arrangement; note that appeals to sustainability or to the quality of the finished steel do not qualify. Section \ref{sec:validation} validates the gate and the extraction.

\begin{figure}[H]
\centering
\colorlet{cRule}{blue!65!black}%
\colorlet{cModel}{violet!75!black}%
\colorlet{cHand}{teal!70!black}%
\setstretch{1}\fontfamily{qhv}\selectfont%
\resizebox{\textwidth}{!}{%
\begin{tikzpicture}[
  stage/.style={rectangle, rounded corners=4pt, draw=#1!60, fill=#1!6,
  very thick, align=center, text width=3.9cm, minimum height=2.6cm,
  inner sep=5pt},
  badge/.style={circle, fill=#1, text=white, font=\bfseries\small,
  inner sep=0pt, minimum size=0.62cm},
  db/.style={cylinder, shape border rotate=90, aspect=0.18, draw=gray!60,
  thick, align=center, text width=2.5cm, minimum height=2.3cm,
  font=\small},
  flow/.style={-{Stealth[length=3mm, width=2.4mm]}, line width=1.3pt, color=gray!60},
  count/.style={font=\footnotesize\bfseries, text=black!80, align=center,
  fill=white, inner sep=1.5pt},
  check/.style={rectangle, rounded corners=3pt, draw=revred!55, fill=revred!5,
  thick, dashed, align=center, text width=3.2cm, minimum height=2.6cm,
  inner sep=5pt, font=\footnotesize},
  probe/.style={-{Stealth[length=2.4mm]}, densely dashed, thick, color=revred!75},
]
\def\dx{6.2}
\node[db, fill=gray!12] (in) at (0,0) {\textbf{CC-News}\\1,163,561,858\\records\\2016–2026};
\node[stage=cRule] (s1) at (\dx,0) {\textbf{Corpus\\preparation}\\[3pt]{\footnotesize\color{black!75} page text split\\by detected language}};
\node[stage=cRule] (s2) at (2*\dx,0) {\textbf{Vocabulary\\screen}\\[3pt]{\footnotesize\color{black!75} steel and scrap terms\\in 49 languages}};
\node[stage=cRule] (s3) at (3*\dx,0) {\textbf{Selection\\gate}\\[3pt]{\footnotesize\color{black!75} three rules decide\\what the model reads}};
\node[stage=cModel] (s4) at (3*\dx,-4.0) {\textbf{Relation\\extraction}\\[3pt]{\footnotesize\color{black!75} one prompt, all languages:\\supplier, buyer, item,\\direction, evidence}};
\node[stage=cModel] (s5) at (2*\dx,-4.0) {\textbf{Entity\\resolution}\\[3pt]{\footnotesize\color{black!75} one English name,\\parent group and\\country per firm}};
\node[stage=cHand] (s6) at (\dx,-4.0) {\textbf{Loop assembly\\and verification}\\[3pt]{\footnotesize\color{black!75} pairing, merging\\duplicates, reading\\the sources}};
\node[db, fill=revred!8, draw=revred!50] (out) at (0,-4.0) {\textbf{44}\\verified\\reciprocal\\loops};
\node[check] (vr) at (4*\dx,0) {\textbf{Recall 72.2\%}\\[2pt]of scrap legs, estimated\\from records the\\gate rejected};
\node[check] (vp) at (4*\dx,-4.0) {\textbf{Precision 94.7\%}\\[2pt]relations read\\against their source};
\foreach \n/\c/\k in {s1/cRule/1, s2/cRule/2, s3/cRule/3, s4/cModel/4, s5/cModel/5, s6/cHand/6}
  \node[badge=\c] at (\n.north west) {\k};
\draw[flow] (in) -- (s1);
\draw[flow] (s1) -- (s2);
\draw[flow] (s2) -- node[count, above=2pt]{10,580,954} (s3);
\draw[flow] (s3) -- node[count, left=3pt]{914,592} (s4);
\draw[flow] (s4) -- node[count, above=2pt]{6,179\\relations} (s5);
\draw[flow] (s5) -- (s6);
\draw[flow] (s6) -- node[count, above=2pt]{44 loops} (out);
\draw[probe] (vr) -- (s3);
\draw[probe] (vp) -- (s4);
\begin{scope}[shift={(1.6,-6.55)}, font=\footnotesize]
  \node[rectangle, rounded corners=2pt, draw=cRule!60, fill=cRule!6, thick, minimum width=0.6cm, minimum height=0.36cm] (l1) at (0,0) {};
  \node[right=3pt of l1] (t1) {rules};
  \node[rectangle, rounded corners=2pt, draw=cModel!60, fill=cModel!6, thick, minimum width=0.6cm, minimum height=0.36cm, right=0.9cm of t1] (l2) {};
  \node[right=3pt of l2] (t2) {language model};
  \node[rectangle, rounded corners=2pt, draw=cHand!60, fill=cHand!6, thick, minimum width=0.6cm, minimum height=0.36cm, right=0.9cm of t2] (l3) {};
  \node[right=3pt of l3] (t3) {automatic pairing and reading by hand};
  \draw[probe] ([xshift=0.9cm]t3.east) -- ++(0.7,0) node[right=3pt, text=black] {validation (Section \ref{sec:validation})};
\end{scope}
\end{tikzpicture}}
\caption{The six stages of the pipeline. Counts are what passes between stages; dashed boxes give the validation of Section \ref{sec:validation} at the stage each measure tests. Appendix \ref{sec:pipeline_details} describes each stage.}
\label{fig:pipeline}
\end{figure}

\subsubsection{Validation Protocol}\label{sec:validation}

We measure precision, the share of returned relations that are correct, and recall, the share of true relations that are returned. All checks use the text the model actually read, which for long records is a text window, not the full article (Appendix \ref{sec:pipeline_details}).

For precision, we checked two samples of returned relations against their source text. The first, of 200 relations, is stratified by language and over-samples the scrap direction, because every loop needs a scrap leg. The second holds the 63 relations in 50 further records, drawn equally from the periods before 2021 and from 2021 onwards. Of these 263 relations, 94.7\% are correct, with a 95\% confidence interval of 91.3\% to 96.8\%.

Recall is harder to measure, because a missed relation leaves no trace in the output. A relation can be lost at the selection gate or by the model, so we check both.

The gate loses every relation in the records it rejects. To find such relations, we ran the same model and prompt on 106,721 rejected records. For relations in the scrap direction, we estimate the gate's recall at 72.2\%, with a 95\% confidence interval of 62.2\% to 81.4\%, so just over a quarter of them never reach the model. None of the lost relations we found would have closed a loop, however, so no loop was among those missed.

To find relations the model overlooks, we read by hand 350 records for which it had returned no relation. They were drawn from three strata, $h = 1, 2, 3$, defined by how many organisations a record names. Stratum $h$ holds $N_h$ records that the model returned empty (145,983, 156,893 and 606,051), and we read $n_h$ of them (110, 90 and 150). Each record read in stratum $h$ therefore stands for $N_h/n_h$ records. If we find $m_h$ missed relations among the records read in stratum $h$, the total number of missed relations is estimated as
\begin{equation}
\hat F = \sum_{h=1}^{3} \frac{N_h}{n_h}\, m_h.
\end{equation}%
The model's recall compares this with $\widehat{TP}$, the estimated number of correct relations the model returned:
\begin{equation}
R_{\text{model}} = \frac{\widehat{TP}}{\widehat{TP} + \hat F}.
\end{equation}%
We found no miss in any stratum ($m_h = 0$), so $\hat F = 0$. Finding none does not prove that the model never misses, so we also bound the miss rate $q_h$, the share of records in stratum $h$ that hold a missed relation. At a miss rate $q_h$, the probability of reading $n_h$ records without meeting a single miss is $(1 - q_h)^{n_h}$. The one-sided 95\% upper bound is the largest miss rate for which this probability is still at least 5\%:
\begin{equation}
q_h^{\max} = 1 - 0.05^{1/n_h}.
\end{equation}%
Across the three strata this bound lies between 2.0\% and 3.3\%.

All labels in these checks were verified by a human reader.

\subsubsection{Statistical Tests}\label{sec:tests}
To test whether a type of loop became more common after 2020, we compare its rate between the two periods. A loop can be observed only between firms that the corpus shows trading steel, so the rate's denominator is the number of such firm pairs, $n_1$ before 2021 and $n_2$ after 2020.

Suppose we observe $k$ loops of a type, $k_1$ of which close in the first period. If the rate had not changed (the null hypothesis), the $k$ loops would behave like a random draw of $k$ of the $n_1+n_2$ pairs, and the number $X$ of them falling in the first period would be hypergeometric. A higher rate after 2020 would instead leave fewer loops before 2021 than chance predicts. The one-sided Fisher exact test therefore takes as its $p$-value the probability that $X$ is no larger than the observed $k_1$:
\begin{equation}
p = \Pr(X \le k_1) = \sum_{j=0}^{k_1} \frac{\binom{n_1}{j}\binom{n_2}{k-j}}{\binom{n_1+n_2}{k}}.
\end{equation}%
Each term is the probability that exactly $j$ loops fall in the first period. When no loop of the type closes before 2021 ($k_1 = 0$), only the first term remains. It is the probability that all $k$ loops fall among the second-period pairs by chance:
\begin{equation}
p = \frac{\binom{n_2}{k}}{\binom{n_1+n_2}{k}} = \prod_{i=0}^{k-1} \frac{n_2 - i}{n_1 + n_2 - i}.
\end{equation}%
Each factor is the chance that the next loop also falls among the second-period pairs, given that the earlier ones did.

Because the quality-motivated rates rest on few or no loops, we give them exact binomial 95\% intervals, one-sided when no loop is observed. Finally, we ask whether quality-motivated loops close later than the others. If the quality labels were unrelated to timing, every reassignment of them among the loops would be equally likely. An exact permutation test therefore compares the observed difference in mean closing year between quality-motivated and other loops with its distribution over all such reassignments.

\subsection{Macro-Level Network Analysis (BACI)}
To quantify the systemic shift, we used the BACI International Trade Database \citep{gaulier2010}. We partitioned material flows into forward (steel products, HS 7208-7229, 7301-7306) and reverse (ferrous scrap, HS 7204) layers. We aggregated data into two periods: $T_1 (2016-2020)$ and $T_2 (2021-2023)$.

\subsubsection{Network Construction and Backboning}
We constructed a weighted directed graph where nodes represent countries and edges represent trade volumes. To extract the significant backbone of the trade network, we applied the disparity filter \citep{serrano2009}. This method filters out non-significant links by comparing the normalized weight of edges against a null model of uniform distribution. We retained only edges with a significance level $\alpha < 0.05$.

\subsubsection{Circularity Metrics}\label{sec:circularity}
We defined the Regional Circularity Index ($C_R$) as the normalized weighted overlap between forward steel exports and reverse scrap imports:
\begin{equation}
  C_R = \frac{\sum_{i,j \in R} \min(P'_{ij}, S'_{ji})}{\sqrt{\left(\sum_{i,j \in R} P'_{ij}\right) \cdot \left(\sum_{i,j \in R} S'_{ji}\right)}}
\end{equation}

Because each overlap term is at most the smaller of the two flows, the numerator cannot exceed the total mass of the smaller layer. The index therefore has a ceiling that depends only on the balance between the two layers:
\begin{equation}
C_R \le k_R^{-1/2}, \qquad k_R = \frac{\max\left(\sum_{i,j \in R} P'_{ij},\ \sum_{i,j \in R} S'_{ji}\right)}{\min\left(\sum_{i,j \in R} P'_{ij},\ \sum_{i,j \in R} S'_{ji}\right)},
\end{equation}%
where $k_R \ge 1$ is the ratio of the larger to the smaller layer mass. A region that trades far more steel than scrap cannot reach a high index, however its flows are arranged.

\subsubsection{Null Model Analysis (Z-Scores)}\label{sec:nullmodel}
The circularity metrics are computed for periods $T_1$ and $T_2$, respectively. 
To determine if the observed shift in circularity was structural or random, we compared the observed values against a random null model (a weighted configuration model, filtered at the same $\alpha$).
More concretely, the flows in the null model are redrawn as
\begin{equation}
  \tilde{w}_{ij} \sim \mathrm{Poisson}\left(s^{out}_i \, s^{in}_j / W\right),
\end{equation}%
with $s^{out}_i$, $s^{in}_j$ total exports and imports and $W$ the layer's total mass.

The evolution of circularity can then be described by the net shift, the observed change relative to the change expected under the null model:
\begin{equation}
  \text{Net Shift} = \left[C_R(T_2) - C_R(T_1)\right] - \left(\mathbb{E}\left[C_R^{rand}(T_2)\right] - \mathbb{E}\left[C_R^{rand}(T_1)\right]\right),
\end{equation}
with $\mathbb{E}$ and $\sigma^{rand}$ taken over 100 draws. %

We calculated the Z-Score as:
\begin{equation}
  Z_{shift} = \frac{\text{Net Shift}}{\sqrt{(\sigma_{T1}^{rand})^2 + (\sigma_{T2}^{rand})^2}}
\end{equation}
For a normally distributed net shift, $Z_{shift} > 1.96$ indicates a statistically significant increase in structural circularity ($p < 0.05$).
It measures distance from randomness rather than effect size and is not comparable across regions; the comparable quantity is rather the excess of $C_R$ over its randomised mean.

\subsubsection{EAF Ratio}
Complementary to the trade network analysis, we assessed the domestic physical capacity for circularity by calculating the EAF ratio for each region, . 
\begin{equation}
  \text{EAF Ratio}_{R,T} = \frac{\sum_{i \in R, t \in T} \text{EAF Production}_{i,t}}{\sum_{i \in R, t \in T} (\text{EAF}_{i,t} + \text{BOF}_{i,t})}
\end{equation}

\section{Results}

\subsection{Structural Embeddedness: Direct Alliances}
Our micro-level news analysis provides evidence of reciprocal loops in both periods and of a change in their stated rationale.

\subsubsection{Temporal Evolution of Closed Loop Alliances}

\begin{figure}[H]
\centering
\includegraphics[width=\textwidth]{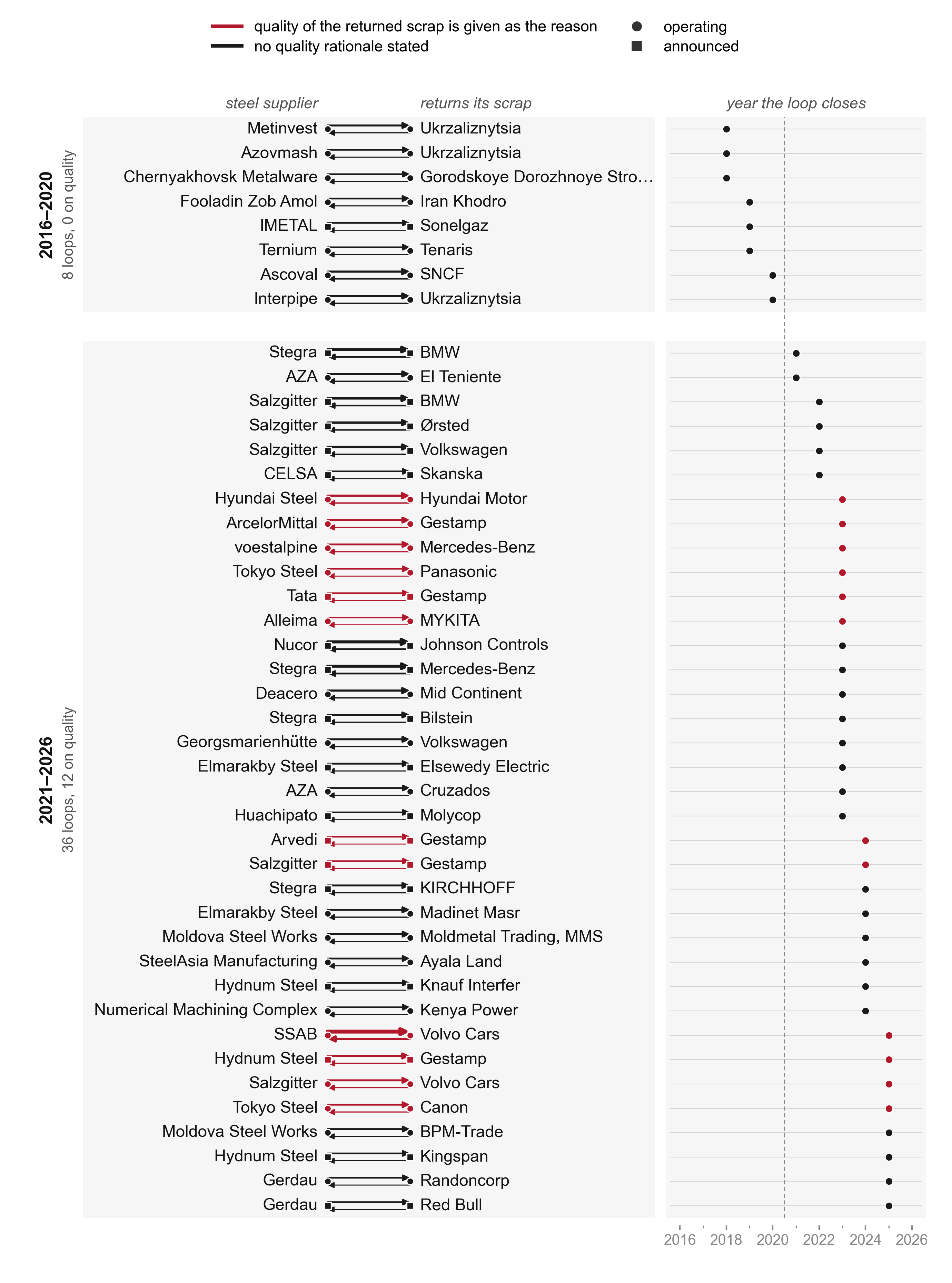}
\caption{The 44 verified loops, placed at the year each closes. Red: quality
rationale; circles: operating; squares: announced.}
\label{fig:closure}
\end{figure}

On the completed corpus the unified pipeline finds 44 verified reciprocal
loops, eight closing before 2021 and thirty-six after (Figure \ref{fig:closure}). Measured against the 527
and 1,619 firm pairs the corpus shows trading steel in the two periods,
reciprocity does not rise significantly (1.52\% to 2.22\%; $p = 0.21$), and end users were already the partner in five of the eight
earlier loops (27 of 36 after). What changes is the stated
rationale (Table \ref{tab:quality}): loops without a quality rationale hold their
rate (1.52\% and 1.48\%, $p = 0.62$), while quality-motivated loops rise from
none to twelve ($p = 0.034$). Quality-motivated loops also
close 1.7 years later than the other types of loop ($p = 0.006$).

\begin{table}[H]
\centering
\caption{Reciprocal loops against the firm pairs the corpus shows trading steel, as count (rate). For quality-motivated loops the 95\% exact binomial interval follows the rate; the first is one-sided. The last row tests for a higher rate after 2020 (Section \ref{sec:tests}).}
\label{tab:quality}
\small
\begin{tabular}{lrrrr}
\toprule
\textbf{Loop closes in} &
\begin{tabular}[b]{@{}r@{}}\textbf{Steel-trading}\\\textbf{pairs}\end{tabular} &
\textbf{All loops} &
\begin{tabular}[b]{@{}r@{}}\textbf{No quality}\\\textbf{rationale}\end{tabular} &
\begin{tabular}[b]{@{}r@{}}\textbf{Quality}\\\textbf{rationale}\end{tabular} \\
\midrule
2016--2020 & 527 & 8 (1.52\%) & 8 (1.52\%) & 0 (0.00\%; $\leq$ 0.57\%) \\
2021--2026 & 1,619 & 36 (2.22\%) & 24 (1.48\%) & 12 (0.74\%; 0.38--1.29\%) \\
\midrule
One-sided Fisher $p$ & & 0.21 & 0.62 & 0.034 \\
\bottomrule
\end{tabular}
\end{table}

\subsubsection{Typology of Detected Loops}
The alliances emerging in this second period reveal a bifurcation in loop maturity. The first category comprises \textit{pre-consumer loops}, representing the optimization of post-industrial waste streams where chemical composition is known and purity is high. This trend is most visible in the European automotive cluster. For instance, Salzgitter has announced closed loops with BMW, Volkswagen, and recently Volvo Cars, where stamping scrap is returned directly to the steelmaker, ensuring the material meets the specific purity requirements of automotive flat steel production. Similarly, in the Nordics, SSAB has forged a steel loop with Volvo Cars. Furthermore, the emergence of dedicated ``green steel'' startups is driving new circular ties, evidenced by H2 Green Steel's explicit scrap-return agreements with BMW and Kirchhoff Automotive, as well as Hydnum Steel's circular pact with Gestamp. Finally, Voestalpine has secured high-grade flows within the DACH region through its alliance with Mercedes-Benz.

More significantly, our analysis detected the emergence of difficult-to-close \textit{post-consumer loops}, signaling the strategic frontier of steel-as-a-service and take-back models. In the energy sector, Salzgitter has extended its reach to post-consumer infrastructure through a novel loop with Ørsted to recycle decommissioned wind turbines. The Nucor--Johnson Controls scrap comes from Johnson Controls' own plants (pre-consumer); Votorantim--Gestamp is not among the 44 loops; post-consumer loops also occur before 2021 (SNCF rails to Ascoval).

\subsection{Macro-Structural Topology: Trade Network Shifts}
\begin{figure}[H]
\centering
\includegraphics[width=\textwidth]{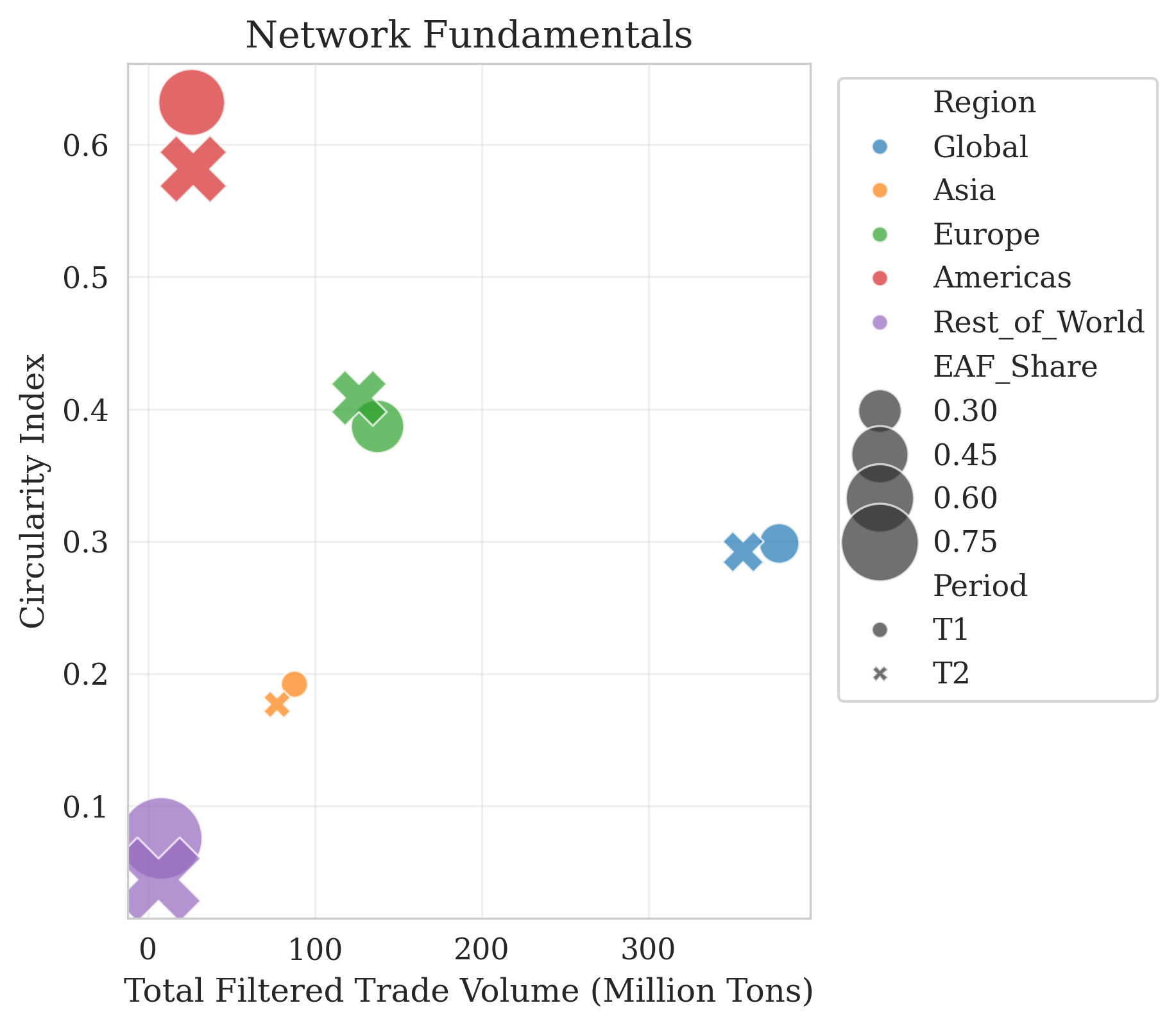}
\caption{Network Fundamentals: Correlation between trade volume, circularity index, and EAF technology share. Bubble size represents EAF share. Note the high circularity of the Americas and Europe relative to the rest of world.}
\label{fig:fundamentals}
\end{figure}

While the news analysis confirms the organizational mechanism, the macroscopic trade network analysis quantifies the systemic impact on global flows. We assessed the fundamental characteristics of the global trade backbone to establish the physical context of circularity (Table \ref{tab:fundamentals} and Figure \ref{fig:fundamentals}).

\begin{table}[htbp]
\centering
\caption{Aggregated Regions Fundamentals (Alpha=0.05)}
\label{tab:fundamentals}
\resizebox{\textwidth}{!}{
\begin{tabular}{llrrrrr}
\toprule
\textbf{Region} & \textbf{Per} & \textbf{Steel Vol (t)} & \textbf{Scrap Vol (t)} & \textbf{Circ Mass (t)} & \textbf{EAF \%} & \textbf{Circ Index} \\
\midrule
Global & T1 & 283,162,534 & 95,328,442 & 49,027,578 & 27.2\% & 0.2984 \\
Global & T2 & 265,083,936 & 91,757,823 & 45,551,055 & 27.9\% & 0.2921 \\
\midrule
Asia & T1 & 76,396,413 & 11,321,149 & 5,644,845 & 17.3\% & 0.1919 \\
Asia & T2 & 68,679,762 & 8,580,935 & 4,288,297 & 17.6\% & 0.1766 \\
\midrule
Europe & T1 & 100,125,286 & 37,460,618 & 23,695,537 & 40.2\% & 0.3869 \\
Europe & T2 & 91,593,680 & 34,803,614 & 23,060,027 & 43.3\% & 0.4084 \\
\midrule
Americas & T1 & 18,081,684 & 7,978,221 & 7,590,440 & 57.7\% & 0.6320 \\
Americas & T2 & 18,538,956 & 8,423,576 & 7,267,137 & 59.0\% & 0.5815 \\
\midrule
Rest\_of\_World & T1 & 7,348,549 & 483,727 & 142,072 & 83.0\% & 0.0754 \\
Rest\_of\_World & T2 & 5,564,528 & 658,606 & 84,415 & 87.2\% & 0.0441 \\
\bottomrule
\end{tabular}
}
\end{table}

\subsubsection{Trade Network Topology ($T_1 \to T_2$)}
\begin{figure}[H]
\centering
\includegraphics[width=\textwidth]{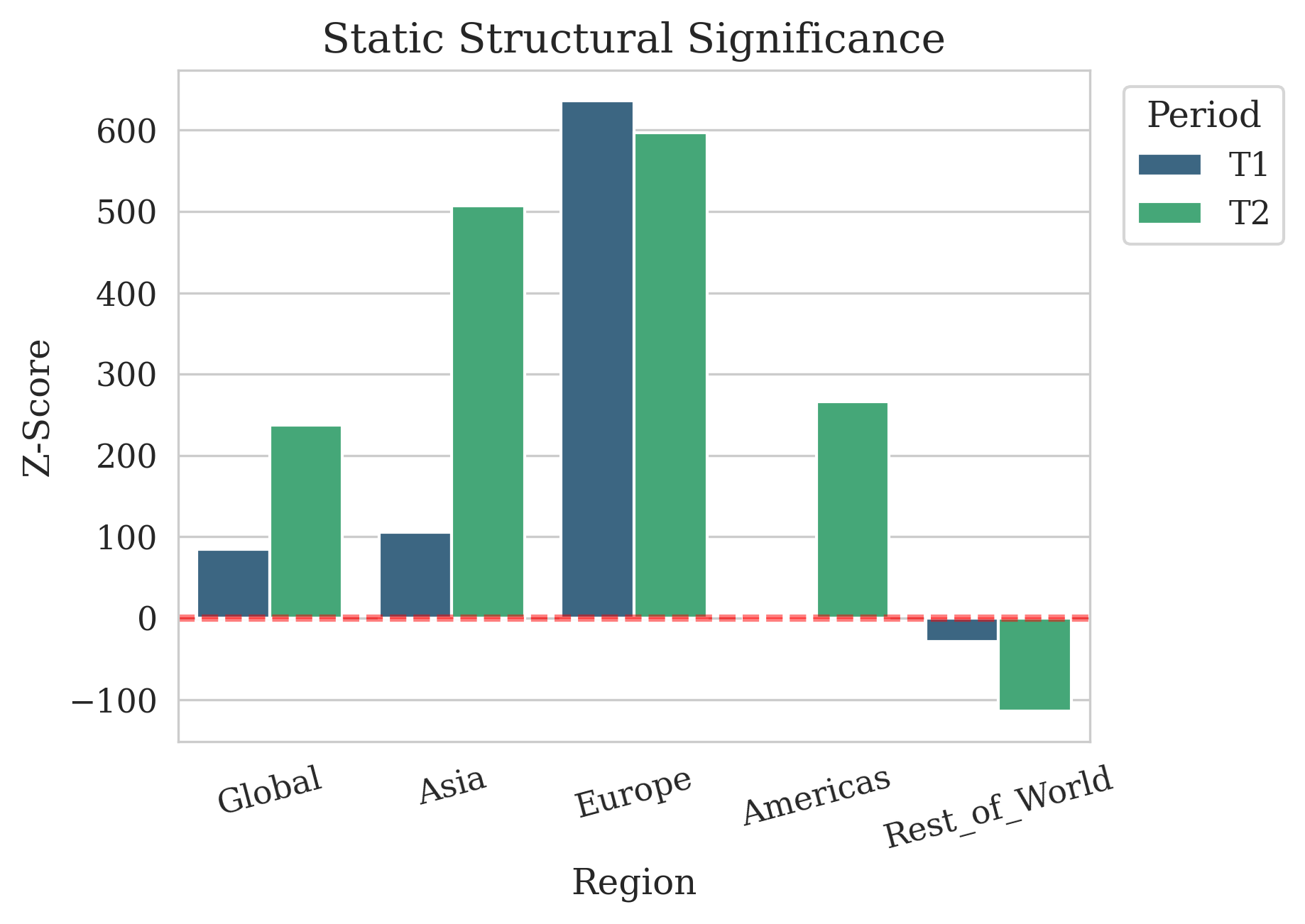}
\caption{Static Structural Significance: Comparison of Z-Scores across regions for T1 and T2. The dashed line represents the significance threshold ($Z=1.96$). Note the shift in the Americas from insignificant (T1) to highly significant (T2) and the absence of such a shift in Europe.}
\label{fig:static_z}
\end{figure}

To ascertain whether these flows signify deliberate circular strategies, we compared the observed circularity against a random null model (Table \ref{tab:static_struct} and Figure \ref{fig:static_z}). The analysis reveals a substantial structural change in the Americas. In the first period ($T_1$), the region's circularity was statistically indistinguishable from a random network ($Z=-0.77$). By $T_2$, despite a modest decline in overall trade volume, the region transitioned to a highly significant circular structure ($Z=+265.72$). In stark contrast, the rest of world (comprising heterogeneous, geographically dispersed and sometimes developing regions in Africa, the Middle East, and Oceania) remains significantly less circular than random ($Z < -100$). 

These $Z$-scores need care, because the index cannot take any value: it is capped by the balance between a region's steel and scrap masses (Section \ref{sec:circularity}). In $T_1$ the Americas already reach 0.632, which is 95\% of their ceiling of 0.664, and the randomised networks sit at almost the same level. The observed network therefore had little room to rise above them, and a $Z$-score near zero does not show that the region's trade lacked organisation. The rest of the world faces the opposite constraint. Its scrap trade is so small next to its steel trade that its index could not exceed 0.26 in $T_1$ or 0.34 in $T_2$, so even a perfectly arranged network there would score low.

A better guide to organisation is how far the observed index lies above its randomised mean, which, unlike $Z$, can be compared across regions (Section \ref{sec:nullmodel}). In $T_2$ the observed index exceeds its randomised mean by 18.1\% globally, 18.5\% in Europe and 25.4\% in the Americas, so trade in these regions is more circular than chance alone would produce.

\begin{table}[htbp]
\centering
\caption{Static Structural Analysis (Real vs Random, Alpha=0.05)}
\label{tab:static_struct}
\resizebox{\textwidth}{!}{
\begin{tabular}{llrrrrr}
\toprule
\textbf{Region} & \textbf{Per} & \textbf{Real Circ} & \textbf{Rand Mean} & \textbf{Rand Std} & \textbf{Z-Score} & \textbf{Sig} \\
\midrule
Global & T1 & 0.2984 & 0.2594 & 0.0005 & 84.50 & *** \\
Global & T2 & 0.2921 & 0.2474 & 0.0002 & 237.25 & *** \\
\midrule
Europe & T1 & 0.3869 & 0.3241 & 0.0001 & 636.11 & *** \\
Europe & T2 & 0.4084 & 0.3446 & 0.0001 & 597.26 & *** \\
\midrule
Americas & T1 & 0.6320 & 0.6358 & 0.0050 & -0.77 & \\
Americas & T2 & 0.5815 & 0.4638 & 0.0004 & 265.72 & *** \\
\midrule
Rest\_of\_World & T1 & 0.0754 & 0.1063 & 0.0011 & -28.95 & *** \\
Rest\_of\_World & T2 & 0.0441 & 0.1247 & 0.0007 & -114.74 & *** \\
\bottomrule
\end{tabular}
}
\end{table}

\subsubsection{Dynamic Structural Evolution}
\begin{figure}[H]
\centering
\includegraphics[width=\textwidth]{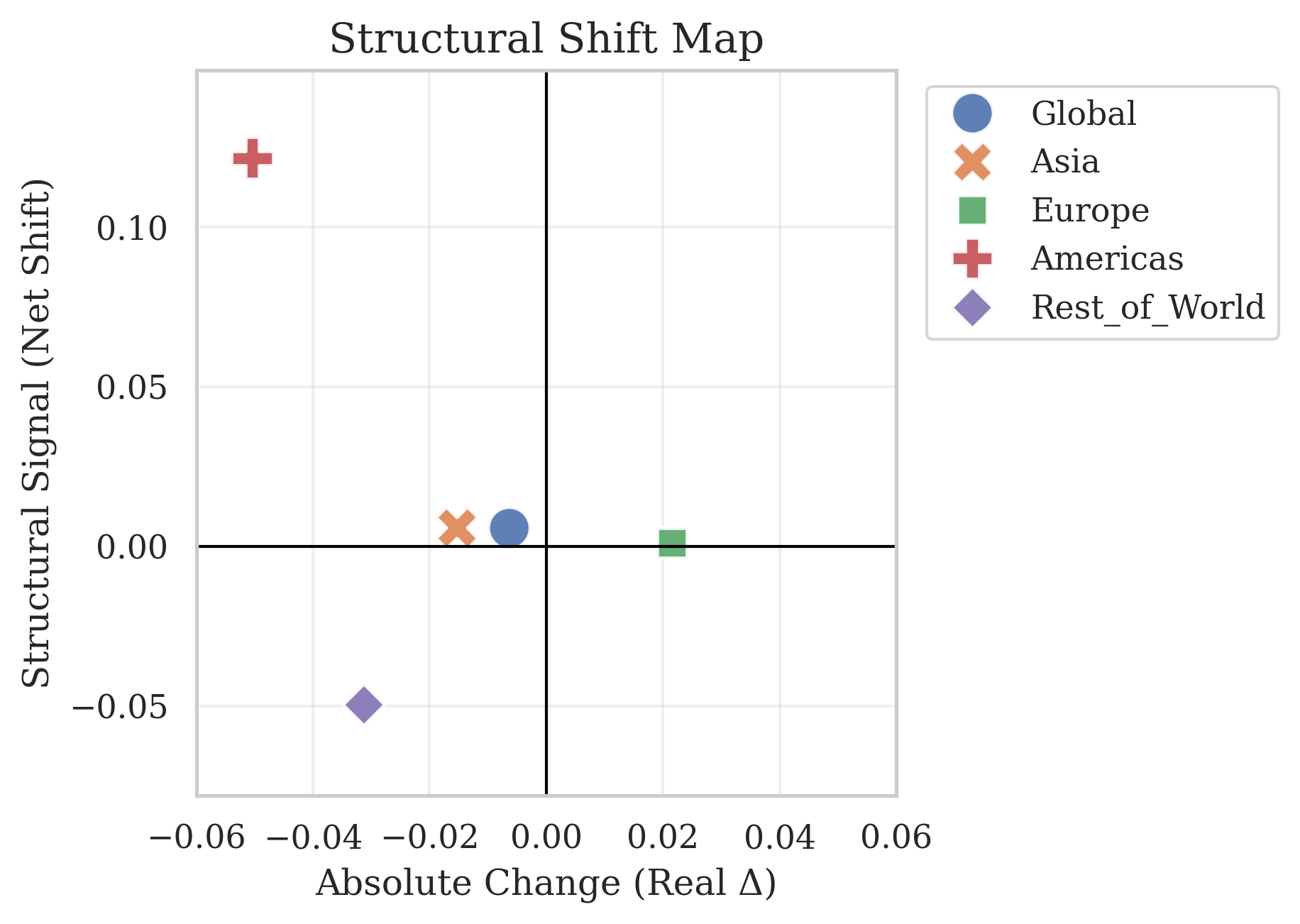}
\caption{Structural Shift Map. The X-axis represents absolute change in circularity ($\Delta_{Real}$), while the Y-axis represents the net structural signal. The Americas appear in a resilient quadrant of negative real change and positive structural net shift.}
\label{fig:shift_map}
\end{figure}

We quantified the evolution of these structures over time using the Z-shift metric (Table \ref{tab:dynamic_shift} and Figure \ref{fig:shift_map}). The Americas show the strongest signal of structural reorganization towards endogenous loop closure ($Z_{shift} = +24.31$). Europe demonstrates a stable, mature circularity ($Z_{shift} = +7.12$). Conversely, the negative shift in the rest of world ($Z_{shift} = -38.81$) . Section \ref{sec:robustness} shows that these shifts are not stable across filter thresholds; we draw no trend from them, and the dating of change rests on Section 4.1.

\begin{table}[htbp]
\centering
\caption{Dynamic Shift Analysis (T1 $\to$ T2)}
\label{tab:dynamic_shift}
\resizebox{\textwidth}{!}{
\begin{tabular}{lrrrrrl}
\toprule
\textbf{Region} & \textbf{Real T1} & \textbf{Real T2} & \textbf{Net Shift} & \textbf{Z(Shift)} & \textbf{Trend} \\
\midrule
Global & 0.2984 & 0.2921 & +0.0057 & +11.36 & INCREASING \\
Asia & 0.1919 & 0.1766 & +0.0057 & +7.35 & INCREASING \\
Europe & 0.3869 & 0.4084 & +0.0010 & +7.12 & INCREASING \\
Americas & 0.6320 & 0.5815 & +0.1215 & +24.31 & INCREASING \\
Rest\_of\_World & 0.0754 & 0.0441 & -0.0497 & -38.81 & DECREASING \\
\bottomrule
\end{tabular}
}
\end{table}

\subsection{Robustness}\label{sec:robustness}
\begin{figure}[H]
\centering
\includegraphics[width=\textwidth]{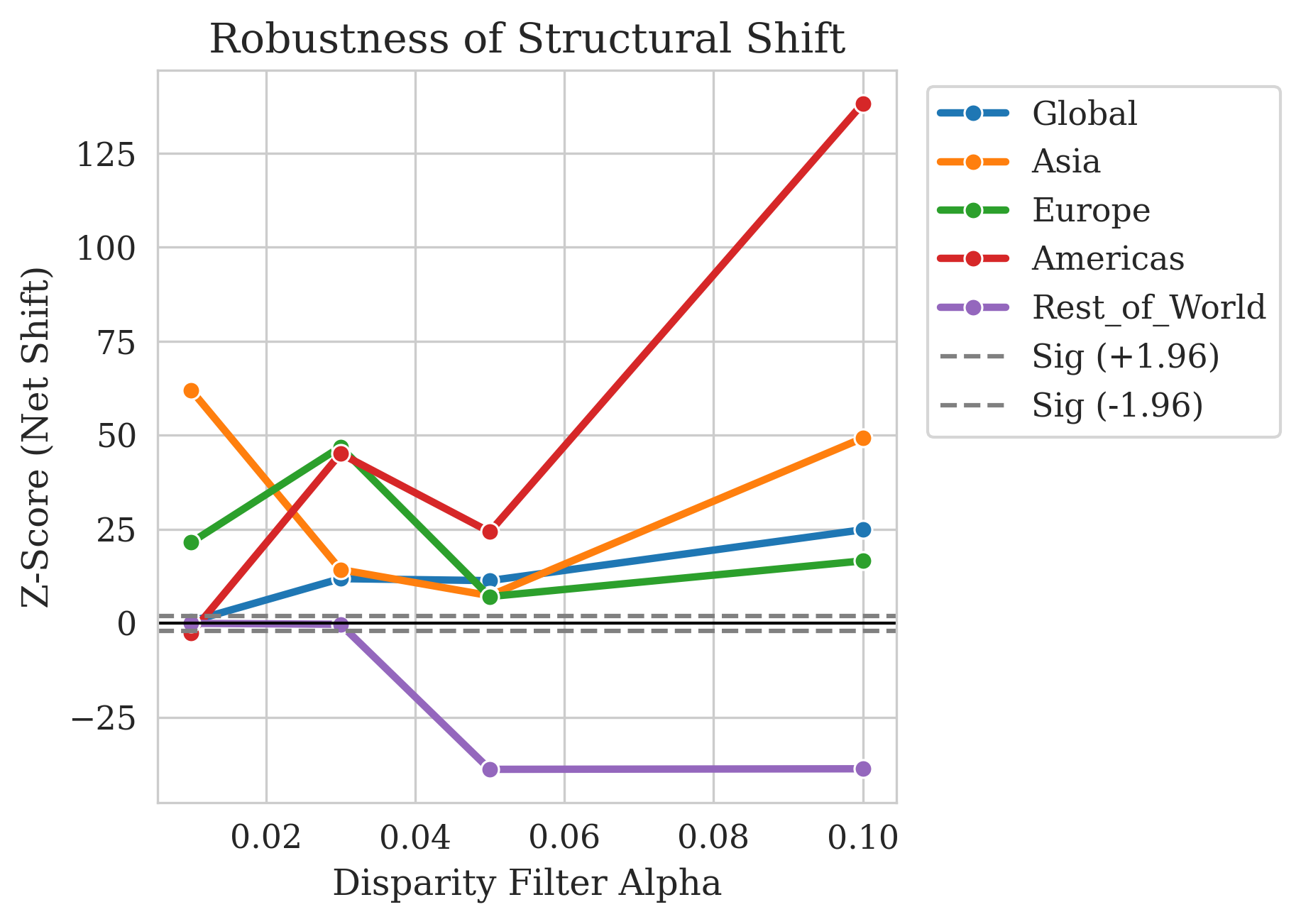}
\caption{Robustness check: structural shift (Z-score) across different disparity filter thresholds ($\alpha$). The vertical axis represents the strength of the structural signal.}
\label{fig:robustness}
\end{figure}

To verify that these topological shifts are not artifacts of specific filtering thresholds employed in the trade network backboning step, we performed a sensitivity analysis (Table \ref{tab:sensitivity} and Figure \ref{fig:robustness}). The strong positive shift in the Americas proves robust across standard thresholds ($\alpha=0.05, 0.10$), reaching its peak at $\alpha=0.10$ ($Z=138.19$). It reverses at $\alpha = 0.01$ ($Z = -2.60$), and the rest of world's shift is zero or near zero at the two strictest thresholds. The convergence of findings, micro-level detection of direct alliances and macro-level statistical shifts in their respective regions (Americas, Europe), provides strong evidence for the bidirectional trade asymmetry hypothesis. This holds for the existence of reciprocal structure, not for its growth.

\begin{table}[htbp]
\centering
\caption{Sensitivity analysis: Z-shift across alphas}
\label{tab:sensitivity}
\begin{tabular}{lrrrr}
\toprule
\textbf{Region} & \textbf{0.01} & \textbf{0.03} & \textbf{0.05} & \textbf{0.10} \\
\midrule
Americas & -2.60 & 45.13 & 24.31 & 138.19 \\
Asia & 61.99 & 14.30 & 7.35 & 49.24 \\
Europe & 21.54 & 46.89 & 7.12 & 16.62 \\
Global & 0.54 & 11.89 & 11.36 & 24.89 \\
Rest\_of\_World & 0.00 & -0.24 & -38.81 & -38.66 \\
\bottomrule
\end{tabular}
\end{table}

\section{Discussion}

We set out to test two hypotheses. Hypothesis~1 holds that, as EAF capacity grows in mature economies, steel exports and scrap imports increasingly overlap within regions. Hypothesis~2 holds that, to reduce uncertainty about scrap quality, producers of high-grade steel form direct alliances with OEMs.

The trade data partly support Hypothesis~1. Relative to chance, the overlap between steel exports and scrap imports increased from the first period to the second, globally and in Asia, Europe and the Americas. The shift is strongest in the Americas ($Z_{shift} = +24.31$), and only the rest of the world moves the other way. We find no clear link to EAF capacity, however. Regions with a higher EAF share are not more circular ($r = -0.11$ and $-0.18$), and a rising EAF share goes with only a slightly larger rise in circularity ($r = +0.21$).

The firm-level data support Hypothesis~2 in a narrower form. Steelmakers and their customers formed reciprocal loops in both periods, but only after 2020 did they give the quality of the returned scrap as the reason. None of the eight loops before 2021 does so, against twelve of the thirty-six after ($p = 0.034$).

\subsection{Theoretical Implications: The De-commoditization of Secondary Resources}

Theoretically, these findings challenge the assumption that circular markets will inevitably evolve towards greater efficiency and standardization. In the linear economy, raw materials like iron ore are quintessential commodities: standardized, fungible, and traded globally on anonymous spot markets. However, our results suggest that the transition to secondary raw materials drives a process of de-commoditization \citep{kopytoff1986cultural}.
Because the quality of post-consumer scrap is opaque and prone to contamination (the market for lemons), the price mechanism alone fails to convey necessary information about material purity. Consequently, the market is not evolving towards the idealized, frictionless trade predicted by standard commoditization theory. Instead, it is reverting to a form of relational governance where trust and provenance replace standardization \citep{dahlmann2019sustainable}. This implies that the circular economy requires a fundamental re-embedding of economic transactions into specific, verified supply relationships. The shift from cost leadership to network sovereignty is therefore not just a strategic choice, but a theoretical necessity for managing the high-entropy nature of waste streams.

\subsection{The Geopolitics of Scrap}
The transition creates a bifurcated global system defined by regional heterogeneity. The region-level reciprocity, observed in the Americas and Europe, is characterized by closure., meaning a significant shift towards reciprocal steel and scrap trade relationships on a country level (Section 4.3). This can be seen as indication that actors prioritize deep, high-dependency relationships to establish resilient, closed-loop ecosystems. However, the trade data is not informative regarding whether this closure is driven by technology or by geopolitics.

In stark contrast, the extractor model observed in the rest of world exemplifies the potential hazards associated with maintaining an open trade posture during a circular transition. Developing regions function as resource peripheries. Without downstream integration, their high EAF adoption accelerates linear extraction (exporting scrap) rather than circular value creation. This creates a dangerous carbon lock-in trap: unable to retain their own high-quality scrap for domestic EAF production, these nations may be incentivized to continue investing in young, long-lived blast furnace capacities to meet infrastructure demand, locking them into high-emission pathways for decades. However, the trade data analysis reveals a more nuanced picture: in $T_2$ the Americas and Europe were net exporters of ferrous scrap (15.2 and 2.3 million tonnes a year) and Asia a net importer (19.6 million).

\subsection{Weaponized Interdependence}
Ultimately, this restructuring is indicative of a broader weaponization of interdependence \citep{farrell2019}, wherein supply chains serve as instruments of statecraft. As nations seek to reduce their linear dependency, the global trade network is bifurcating into distinct, regionally concentrated cycles.

China further fragments this landscape through its regulatory decoupling. In the wake of the ``National Sword'' policy, China effectively decoupled from the global scrap trade to cultivate a strictly domestic circular ecosystem. While this strategy does not manifest as reciprocal cross-border trade, it represents the ultimate form of sovereign circularity: autarky. This suggests that the mechanism of closure can manifest as regional integration (the Americas/Europe) or national protectionism (China), both serving the same goal of security. In a world of volatile geopolitics, owning the physical molecules of the material cycle is increasingly viewed as a form of national security.

Another potential explanation for the region-level reciprocity of steel and scrap flows simply lies in scrap logistics. Particularly for land-locked countries where railway is the major modality for scrap transport, the formation of closed-loop alliances might be regionally constrained by transport distance. So while logistical constraints certainly play a role for specific countries, they are unlikely to be the sole factor, as evidenced by the Chinese National Sword Policy and corroborated by our findings of substantial heterogeneity across manufacturers even within regions and the highly non-random network structure we observe.

\subsection{Implications for Renewable Energy Production}
A noteworthy finding of our micro-level analysis is the intersection of the steel cycle with the energy transition, exemplified by the Salzgitter-Ørsted alliance to recycle wind turbines. This represents a double loop: the green transition requires massive steel inputs for renewable infrastructure (wind towers), which eventually become the high-grade scrap needed to produce future green steel. This nexus suggests that the seminal steel scrap age is closely linked to the lifecycle of energy infrastructure. Regions that led the early deployment of renewables (like Northern Europe) may now be the first to harvest this energy scrap, providing them with a secondary advantage in the transition towards green steel.

\subsection{Managerial and Policy Implications}
From a managerial perspective, these findings signal a reduced role of the global spot market for strategic raw materials. The paradigm of competition through supply chains \citep{sirkin2005} might evolve into a competition for circular control. Consequently, ferrous scrap must be reclassified from a generic leverage item to a strategic raw material.

The viability of steel-as-a-service models has historically been hindered by high transaction costs. However, our identification of post-consumer loops, together with the significant increase in quality-motivated loops, suggests that the existential threat of quality contamination may be lowering these barriers.

For policymakers in developing regions, domestic scrap retention is critical. Incentives must shift towards domestic recycling infrastructure. In the Global North, regulatory frameworks should support mechanisms to trace provenance (such as product passports) to facilitate these direct take-back schemes, potentially granting end-of-waste status to verified closed-loop materials to reduce regulatory friction.

\subsection{Limitations}
Our study is subject to specific limitations. First, the BACI database covers international trade only, meaning substantial domestic circularity (e.g., within the US or China) remains imperceptible to our macroscopic trade network metrics. Second, the null model cannot fully isolate specific policy interventions (e.g., CBAM) from organic market shifts. Finally, a key limitation is that trade data analysis cannot distinguish between scrap qualities and pre/post consumer differences. However, it provides quantitative evidence that direct reciprocal ties exist on a geopolitical level between countries, triangulating the qualitative findings of our micro-level analysis.

Regarding the micro-level analysis, while our validation protocol confirms a high degree of precision (94.7\% on the corpus, Section \ref{sec:validation}), the recall metrics suggest that the automated pipeline likely underestimates the total number of circular alliances. Despite the extensive scale of the Common Crawl corpus, it may not capture every niche industrial publication or paywalled contract announcement, leading to potential gaps in coverage. Furthermore, the linguistic complexity of steel supply contracts, often obfuscated by broad commercial language, means that the identified cycles likely represent a conservative lower bound of the actual extent of circular integration. Our pipeline functions as a high-confidence filter, prioritizing the suppression of false positives to ensure topological validity over the exhaustive retrieval of every potential signal.

\section{Conclusion}
The steel scrap age is not leading to a globalized market for scrap, but to a regionalized, organizationally integrated landscape. The trade data show reciprocal structure in both periods but not its growth; after 2020 reciprocal loops began to be justified by the quality of the returned scrap, in twelve of thirty-six against none of eight before. Confronted with the failure of spot markets to signal quality, steelmakers are actively constructing direct, verified loops. This structural reorganization may define the industrial geography of the low-carbon transition. As nations and firms endeavor to decouple from carbon-intensive supply chains, the global trade network may shift from an efficiency-driven logic of globalization to a resilience-driven logic of regional circularity \citep{witt2019}.

\newpage
\bibliography{references}

\newpage

\appendix

\section{Corpus Examples}
\label{sec:corpus_examples}

Records are given in English; translations are ours. For each loop the first source of each leg is shown; the full source lists are in the replication files.

\subsection{A quality-motivated loop: SSAB and Volvo Cars}
\textit{Steel leg} (nieuws.nl, 2021, Dutch): Volvo Cars will examine building cars from fossil-free steel, and SSAB will make the steel. \textit{Scrap leg} (SSAB interim report on finanznachrichten.de, 2025, English): SSAB reports the ``purchase of high-quality scrap steel from Volvo Cars' production''. The loop closes in 2025 and is operating; its sources give a property of the returned scrap, so it is coded quality-motivated.

\subsection{A post-consumer loop before 2021: Ascoval and SNCF}
\textit{Steel leg} (Reuters on boursorama.com, 2019, French): Ascoval agrees to supply SNCF with 140,000 tonnes of steel a year for four years for rail making. \textit{Scrap leg} (boursier.com, 2020, French): SNCF's used rails are remelted by Ascoval. The loop closes in 2020 and is operating; no source gives a reason tied to the scrap, so it is not quality-motivated.

\subsection{Appliance steel returned to the mill: Tokyo Steel and Panasonic}
\textit{Both legs} (Panasonic press release on jiji.com, 2023, Japanese): steel that Tokyo Steel recycled from used home appliances is used for the Panasonic Group pavilion at Expo 2025, and after the Expo the steel goes back to Tokyo Steel. The loop closes in 2023 and is coded quality-motivated.

\subsection{A loop with no English source: Iran Khodro and Fooladin Zob Amol}
\textit{Scrap leg} (irna.ir, 2019, Persian): under its commitment Iran Khodro was to deliver four thousand tonnes of scrap iron a month to the Fooladin Zob Amol foundry. \textit{Steel leg} (irna.ir, 2019, Persian): about 90\% of the foundry's cast-iron and steel parts go to Iran Khodro and its parts supplier SAPCO. The loop closes in 2019.

\subsection{Relations the validation rejected}
(dw.com, 2023, Ukrainian): Zaporizhstal handed steel for the emblem on the Motherland Monument to the contractor Osnova free of charge; the model returned a supply, but a transfer without payment is excluded. (teslarati.com, 2020, English): Redwood Materials reclaims battery-cell scrap from Panasonic's production at Tesla's Nevada plant; the model returned a scrap relation, but the scrap is not ferrous.

\subsection{A screened record with no relation}
(theenterpriseleader.com, 2023, English): an investor's change in its Norfolk Southern shareholding. The company profile lists steel and scrap metals among the freight carried, so the record passes the screen and the gate; the model correctly returns no relation.

\section{Pipeline Details}
\label{sec:pipeline_details}

This appendix gives the technical content of each stage in Figure \ref{fig:pipeline}. The replication files contain the code, together with the complete steel and scrap term lists for all 49 languages, the unambiguous industry terms and closed-loop phrases used by the selection gate, and the terms the audit excluded.

\subsection{Corpus preparation}
The CC-News archive is read from its monthly WARC files. For each archived page, the text of every paragraph longer than five words is taken from the HTML, and its language is detected with \textit{langdetect}; a paragraph is kept when the most probable language has probability above 0.95. Paragraphs of the same language are joined, so a page yields one record per language it contains. A record carries the page URL, its text, its language and the year and month of the archive file, which dates it.

\subsection{Vocabulary screen}
Each language has its own list of steel and scrap terms. A record is filed under a language by its script first and its detected label second, because language detection often mislabels Chinese, Japanese and Korean text: a record whose opening is not ASCII and contains kana, hangul or Chinese characters is filed under Japanese, Korean or Chinese. The lower-cased text is then matched against that language's terms, with word boundaries on both sides except in Chinese, Japanese, Korean and Thai, which have none. A record is retained when at least one term matches and its length lies between 60 and 30,000 characters and between 30 and 3,000 words. The screen stores the matched terms with each record, so later stages need not reread the archive.

\subsection{Selection gate}
The gate applies three rules. (i) In the 37 languages whose retained records are few enough to be read in full, every record is selected. In the other twelve languages, terms that an audit showed are not about metal in practice are first removed, and a record is selected when (ii) it carries both a steel and a scrap term, or (iii) it carries an unambiguous industry term (in English, for example, \textit{steelworks} or \textit{ferrous scrap}) together with a scrap mention, or it contains a closed-loop phrase (in English, for example, \textit{closed loop}, \textit{take-back} or \textit{circular economy}). Scrap mentions at this stage are matched with a looser pattern that also finds a scrap word at the end of a compound, such as German \textit{Presswerkschrott}; the extraction window uses the same pattern, so a record selected for a scrap mention is shown that mention.

\subsection{Relation extraction}
The language model is \texttt{gpt-5.6-sol} with medium reasoning effort. Records are sent sixteen at a time under one English system prompt for every language. The prompt counts a relation only when the text describes an actual transaction, supply agreement or take-back arrangement between two named companies involving steel or ferrous scrap, and it excludes share prices, ratings, lawsuits, plant closures, capacity, policy, conferences, donations and crime reports. Each relation is returned as supplier, buyer, item, direction (steel or scrap, the material flowing from supplier to buyer), a flag set when the text says the pair also trades in the reverse direction, and a quotation of at most twelve words from the text. A record longer than 3,200 characters is shown as its first 1,200 characters plus 380 characters on either side of its first and last scrap mention, because the reverse leg of a loop is often a clause late in an announcement.

\subsection{Entity resolution}
The model returns each firm as the record names it, so one firm appears under several spellings and scripts. Resolution does not reread the records: the distinct firm names are sent to the same model in alphabetical blocks of forty, so that variants of one firm tend to be seen together. For each name the model returns the firm's usual English name without legal form, its ultimate parent group, its home country and its type (steelmaker, scrap merchant, steel user or other), and it is instructed never to merge two genuinely different companies.

\subsection{Loop assembly and verification}
A candidate loop is a pair of firms with a steel relation in one direction and a scrap relation in the other. Relations are matched on the resolved name or on the parent group, in either orientation, since a relation does not fix which side is written first. Candidates that describe the same arrangement under different names are merged. Each candidate was then read against its sources: loops that are not steel--scrap exchanges, such as aluminium, were removed. Each surviving loop is dated by the year it closes and coded for a quality rationale as defined in Section \ref{sec:pipeline}, and as operating or announced (Figure \ref{fig:closure}).

\end{document}